\documentstyle[aps,prb,preprint]{revtex}

\begin{document}
\draft
\title{Kinetics of exciton photoluminescence  in type-II 
semiconductor superlattices} 

\author{L.~S.~Braginsky\thanks{Electronic address: 
brag@isp.nsc.ru},  M.~Yu.~Zaharov, A.~M.~Gilinsky,
V.~V.~Preobrazhenskii, M.~A.~Putyato, and K.~S.~Zhuravlev}
\address{Institute of Semiconductor Physics, 630090, 
Novosibirsk, Russia}

\date{\today}
\maketitle
\begin{abstract}
The  exciton decay rate  at a rough 
interface in type-II semiconductor superlattices is 
investigated. It is shown that the possibility of recombination 
of indirect excitons at a plane interface essentially affects 
kinetics of the exciton photoluminescence at a rough interface. 
This happens because of  strong correlation between the 
exciton recombination at the plane interface and at the 
roughness.  Expressions that relate the parameters of the 
luminescence kinetics with statistical characteristics of the 
rough interface  are obtained.  The mean height and 
length of roughnesses in GaAs/AlAs superlattices are estimated 
from the experimental data.  
\end{abstract} 
\narrowtext 
\section{Introduction}
GaAs/AlAs type-II superlattices are the subject of extensive 
investigations in the recent decade. Electrons and holes are 
separated in these structures: holes are confined in the 
$\Gamma$ valley of GaAs, whereas electrons are   in $X$ valleys 
of AlAs. Changing the width of AlAs layer during the growth 
of the structure, it is possible to confine the electrons 
either in the $X_z$ valley ($X$ valley that is directed 
along the structure axis [001]) or in the $X_{xy}$ valley 
($X$ valley that is directed along the GaAs/AlAs interface: 
[100] or [010]).  The excitons in such structures are indirect 
in both real and momentum spaces. 

Kinetics of the exciton luminescence is 
investigated by the time-resolved method.
The theory by Klein {\it et al. } \cite{Klein} is usually used 
to explain the results of such experiments.  The theory 
has been developed to consider the no-phonon 
radiative decay rates of indirect excitons in  alloy 
semiconductors (e.g., Ga$_{1-x}$Al$_x$As).  The recombination of 
indirect excitons occurs because of intervalley scattering of 
electrons at the potential fluctuations caused by the 
compositional disorder. These short-range scatterers are 
necessary to compensate the large momentum of the electron in 
the $X$ valley. The nonexponential dependence of the decay rate 
has been obtained 
\begin{equation} 
\label{1} 
I(t)\propto e^{-w_0t}(1+2w_rt)^{-3/2}.  
\end{equation} 
Where the value $w_r$ is connected with the compositional 
disorder. The exponential factor has been included in 
Eq.~(\ref{1}) to consider  different nonstochastic processes of 
the exciton recombination (e.g., the phonon-assisted 
recombination).  This is possible only in  the absence of 
correlation between stochastic and nonstochastic processes.

The possibility to apply the theory \cite{Klein} for 
superlattices has been discussed by F.~Minami 
{\it et al.}.~\cite{Minami} Authors suppose the short-range 
scatterers are distributed along the plane boundary.  This 
assumption justifies the application of Eq.~(\ref{1}) for 
superlattices; however, it does not allow to relate the 
parameter $w_r$ with characteristics of the rough interface, 
e.g., the mean height and length of roughnesses. I.  Krivorotov 
{\it et al.}\cite{Krivirotov} have showed that nonradiative 
decay due to exciton trapping by interfacial defects also 
results in nonexponential factor in Eq.~(\ref{1}).  
Nevertheless, Eq.~(\ref{1}), wherein the parameters $w_r$ and 
$w_0$ are considered as trial, is commonly used for explanation 
of the experimental results.\cite{Nagao}

It should be noted that  roughnesses are not necessary  for 
the recombination of $X_z$ excitons. Their recombination occurs 
even at a plain interface where the normal component of the 
electron momentum relaxes. This  important 
point also distinguishes the exciton recombination in 
superlattices.  The process, however, can not be taken into 
account by a simple exponential factor. Indeed, the wave function 
of the electron at a rough interface is the 
sum of its regular and diffuse components. The first one exists 
at a plane interface, whereas the latter is due to the 
roughnesses. For this reason the crossed terms arise in 
the interband matrix element; so that the probability of the  
exciton recombination, which is determined by the squared module 
of this matrix element, is no longer   a simple sum of the 
probabilities of the  recombination at the plane 
interface and at the roughnesses.  This correlation leads to a 
more complicated relation than the simple exponential factor in 
Eq.~(\ref{1}).

In this paper we consider a more realistic model of the rough 
interface. We show that Eq.~(\ref{1}) holds for 
the decay rate of  $X_{xy}$ excitons and   relate 
the $w_r$ value with parameters of the rough interface. We 
determine the decay rate of  $X_z$ excitons. In particular,   
it is found that this value at  large times behaves roughly as 
$I(t)\propto \exp (-w_0t)/t$, rather than $I(t)\propto \exp 
(-w_0t)/t^{3/2}$ as it is predicted by Eq.~(\ref{1}). 

Our experiments on the GaAs/AlAs type-II superlattices confirm 
these results. We use the experimental data for the radiative 
decay rates to estimate the parameters of the rough interface. 
The mean height of roughnesses was found to be close to  
the lattice constant, whereas their mean length 
is about $50\,\AA$. 

\section{Radiative decay rates of indirect excitons in 
superlattices. Theory}

Let $z=0$ be the interface between GaAs ($-d_1<z<0$) and AlAs 
($0<z<d_2$), and  $\bbox{\rho}$ be the vector in the $XY$ plane.
We consider the exiton recombination at the interface 
and write the exciton wave function as 
follows:\cite{Ivchenko}
\[
 \phi({\bf r}_e,{\bf r}_h)=f_e(z_e)f_h(z_h)
G(\bbox{\rho}_e-\bbox{\rho}_h,z_e,z_h).
\]
Where ${\bf r}_e=\{\bbox{\rho}_e,\ z_e\}$ and ${\bf 
r}_h=\{\bbox{\rho}_h,\ z_h\}$ are coordinates of  the electron 
and the hole,  $f_e(z_e)$ and $f_h(z_h)$ are their wave 
functions  in the absence of Coulomb interaction; the 
function $G$  takes into account this interaction.  The 
probability for the exciton recombination is proportional to 
$G^2({\bf 0})$
[$G({\bf 0})\equiv G(\bbox{\rho}_e=\bbox{\rho}_h,z_e=z_h=0)$] 
and the squared module of the matrix element 
\begin{equation} 
\label{2} 
{\cal P}= \int 
f_e(z)\bbox{\nabla}f_h(z)\,dz\,d^2\bbox{\rho}. 
\end{equation}
 
The  functions $f_e(z_e)$ and $f_h(z_h)$ can be expressed via 
envelope wave functions of the electron and the hole in the 
conduction and valence bands of GaAs and AlAs. To determine the 
envelopes, the appropriate boundary conditions at the GaAs/AlAs 
interface should be imposed. The roughness of the interface 
has an influence on these boundary conditions and, therefore, 
affects the envelopes. We shall consider the rough interface 
where the mean height of roughnesses is small in comparison 
with the electron wavelength (or the exciton size).  This 
allows us to use the boundary conditions at the rough 
interface\cite{PhysicaE} to consider an influence of 
roughnesses on the exciton recombination. 

\subsection{Boundary conditions for the envelope wave functions 
at a GaAs/AlAs interface}
\subsubsection{Boundary conditions at a plane interface}
In general, the boundary conditions for the electron 
envelopes can be written as follows:
\begin{equation}
\label{3}
\left(
\begin{array}{l}
\Psi_\Gamma^r\\
\Psi_\Gamma^{r'}\\
\Psi_{Xxy}^r\\
\Psi_{Xxy}^{r'}\\
\Psi_{Xz}^r\\
\Psi_{Xz}^{r'}
\end{array}
\right)
=\tilde{T}
\left(
\begin{array}{l}
\Psi_\Gamma^l\\
\Psi_\Gamma^{l'}\\
\Psi_{Xxy}^l\\
\Psi_{Xxy}^{l'}\\
\Psi_{Xz}^l\\
\Psi_{Xz}^{l'}
\end{array}
\right).
\end{equation}
Where  $\Psi_{\Gamma,Xxy,Xz}^{l,r}$ are the envelopes which 
correspond to the $\Gamma$, $X_{xy}$, and $X_z$ valleys of GaAs 
and AlAs; $\Psi_{\Gamma,Xxy,Xz}^{l,r'}\equiv \partial 
\Psi_{\Gamma,Xxy,Xz}^{l,r}/\partial z$ are their normal 
derivatives.  The elements $\tilde{t}_{ik}$ of the $6\times 6$ 
matrix $\tilde{T}$ are determined by the interface structure. 
They are independent of the electron energy. For the GaAs/AlGaAs 
interface they have been calculated by Ando {\it et 
al.}.\cite{Ando} 

We shall consider the particular cases of $X_z$ and $X_{xy}$ 
excitons.   This allows us to simplify Eq.~(\ref{3}).  
First, we omit mixing between $X_z$ and $X_{xy}$ valleys.  
Second, the energy position of $\Gamma$ minimum in AlAs is 
considerably higher than that of $X$ minimum.  For this reason 
the wave function $\Psi_\Gamma^r$ decays rapidly apart of the 
interface.  We have $\Psi_\Gamma^r\propto \exp(-\gamma^r z),\ 
\Psi_\Gamma^{r'}=-\gamma^r\Psi_\Gamma^{r}$, where 
$\gamma^r=\sqrt{2m_{\Gamma}^r(E_\Gamma-\varepsilon_e)}$ (here 
$m_{\Gamma}^r$ is  effective mass in the $\Gamma$ valley of 
AlAs, $\varepsilon_e\approx E_X$ is the electron energy, 
$E_\Gamma$ and $E_X$ are energies of bottoms of the $\Gamma$ and 
$X$ valleys) can be considered as independent of the electron 
energy. By  eliminating of $\Psi_\Gamma^{r}$ from Eq.~(\ref{3}),  
for $X_z$ electrons we find:  
\begin{mathletters} 
\label{bcplane} 
\begin{equation}
\label{4}
\left\{
\begin{array}{l}
\Psi_{Xz}^r=\Psi_{Xz}^l,\\
\Psi_{Xz}^{r'}= 
t^z_{41}\Psi_\Gamma^{l}+t^z_{44}\Psi_{Xz}^{l'},\\ 
\Psi_\Gamma^l+t^z_{12}\Psi_\Gamma^{l'}+ t^z_{13}\Psi_{Xz}^l=0.
\end{array}
\right.
\end{equation}
Where $t^z_{44}\approx m_{Xl}^r/m_{Xl}^l\approx 1$,  this value 
takes into account the difference of longitudinal effective 
masses in the $X$ valleys of AlAs and GaAs;  $t^z_{41}=t_{\Gamma 
X} m^r_{Xl}/(m_ea)$, $t^z_{12}=m^r_\Gamma/(m^l_\Gamma\gamma^r)$, 
$t^z_{13}=t_{\Gamma X}m^l_\Gamma/(m_ea\gamma^r)\ll 1$; 
$t_{\Gamma X}\approx 1$ is the parameter of $\Gamma$--$X$ 
mixing. Other elements of the ${t}^z_{ik}$ matrix   are small; 
this is the result of numerical calculations.\cite{Ando} 

Note that the band states in the $X$ valley   result from 
interaction of two close-lying bands: lower $X_1$ and upper 
$X_3$; meanwhile only the $X_3$ states mix effectively with 
$\Gamma$ states.  This means that  $t_{\Gamma X}\approx 1$ is 
the upper estimation of $\Gamma$--$X$ mixing.

It is sufficient to consider only $X$ valleys of each 
contacted material when $X_{xy}$ electrons are investigated. 
Assuming $\Psi_\Gamma^{l'}=\gamma^l\Psi_\Gamma^{l}$, where 
$\gamma^l\sim 2\pi/a$ ($a$ is the lattice constant), from 
Eq.~(\ref{3}) we find: 
\begin{eqnarray} 
\label{5} 
&&\Psi_{Xxy}^r=t^{xy}_{11}\Psi_{Xxy}^l+t^{xy}_{12}\Psi_{Xxy}^{l'},\\
&&\Psi_{Xxy}^{r'}=t^{xy}_{21}\Psi_{Xxy}^l+t^{xy}_{22}\Psi_{Xxy}^{l'}.
\nonumber 
\end{eqnarray}
Where $|t^{xy}_{12}|\ll 1,\ |t^{xy}_{21}|\ll a^{-1},\ 
|t^{xy}_{11}|\approx 1$, and $t^{xy}_{22}\approx 
m_{Xt}^r/m_{Xt}^l\approx 1$; $m_{Xt}^r$ and $m_{Xt}^l$ are  the
transversal effective masses of AlAs and GaAs.

The bands of the light and heavy holes  are splitted due 
to the size quantization.  
This allows us to consider only the heavy holes in each material 
and write the boundary conditions for them as follows: 
\begin{eqnarray} 
\label{6} 
&&\Psi_h^r=t^h_{11}\Psi_h^l+t^h_{12}\Psi_h^{l'},\\
&&\Psi_h^{r'}=t^h_{21}\Psi_h^l+t^h_{22}\Psi_h^{l'}.
\nonumber 
\end{eqnarray}
Where $\Psi_h^{l,r}$ are the envelopes for the heavy holes in 
each material. 
\end{mathletters}

\subsubsection{Boundary conditions at a rough interface}
We shall consider the model of a rough interface that is 
presented on Fig.~1.  This model is in  agreement with optical 
\cite{Lurssen} and structural \cite{Bechstedt}  investigations 
of GaAs/AlAs interface.  The interface looks like an array of 
the plane areas of the same crystallographic orientation. The 
random function $z=\xi(\bbox{\rho})$ of the  coordinates in the 
$XY$ plane determines the positions of these areas relative to 
$z=0$. 

We  assume the average 
height of roughnesses $h$ to be small in comparison with 
the electron wavelength. Then it is possible to describe the 
rough interface by means of the correlation function $W(\bbox 
{\rho '}, \bbox{\rho ''})=\overline{\xi(\bbox{\rho '})\xi(\bbox{\rho 
''})}$.  For the homogeneous  rough interface $W(\bbox{\rho 
'}, \bbox{\rho ''})=W(\bbox{\rho '}-\bbox{\rho ''})$, i.e.,   
the correlation function is the function of one variable:  
$\bbox{\rho}=\bbox{\rho '}-\bbox{\rho ''}$.  There are two 
parameters that are most important when the statistical 
properties of a rough interface are considered:  $h^2=W(0)$ and 
the correlation length $l$ --- the mean attenuation length of 
the correlation function.  In our model the correlation length 
$l$ can be associated with the mean size of the plane area. 

The special form of the rough interface (Fig.~1) allows  us to 
apply the boundary conditions (\ref{bcplane}), which are 
applicable at a plane interface, at each plane $z=\xi$. 
The inequality $|\xi\Psi'|\sim h/\lambda\ll 1$ ($\lambda$ is the 
electron wavelength) allows to rewrite these boundary conditions 
at a plane $z=0$.  After some algebra  we obtain:
\begin{mathletters} 
\label{bcrough}
\begin{eqnarray}
\label{7}
&&\Psi_{Xz}^r=-t^z_{41}\eta(\xi)\xi(\bbox{\rho})\Psi_\Gamma^l
+\Psi_{Xz}^l+
(1-t^z_{44})\xi(\bbox{\rho})\Psi_{Xz}^{l'},\nonumber\\
&&\Psi_{Xz}^{r'}=t^z_{41}\eta(\xi)\Psi_\Gamma^l
+t^z_{41}\eta(\xi)\xi(\bbox{\rho})\Psi_\Gamma^{l'}
t^z_{44}\Psi_{Xz}^{l'},\\
&&\Psi_\Gamma^l+[t^z_{12}+\xi(\bbox{\rho})]\Psi_\Gamma^{l'}
+t^z_{13}\eta^*(\xi)\Psi_{Xz}^l
+t^z_{13}\eta^*(\xi)\xi(\bbox{\rho})\Psi_{Xz}^{l'},
\nonumber 
\end{eqnarray}
for  electrons in the $X_z$ valley;
\begin{eqnarray}
\label{8}
&&\Psi_{Xxy}^r=\Psi_{Xxy}^l+(1-t^{xy}_{22})\xi(\bbox{\rho})\Psi_{Xxy}^{l'},\\
&&\Psi_{Xxy}^{r'}=t^{xy}_{21}\Psi_{Xxy}+t^{xy}_{22}\Psi_{Xxy}^{l'},
\nonumber 
\end{eqnarray}
for  electrons in the $X_{xy}$ valley; and
\begin{eqnarray}
\label{9}
&&\Psi_h^r=\Psi_h^l+(1-t^h_{22})\xi(\bbox{\rho})\Psi_h^{l'},\\
&&\Psi_h^{r'}=t^h_{21}\Psi_h+t^h_{22}\Psi_h^{l'},
\nonumber 
\end{eqnarray}
for the holes. Factor $\eta(\xi)=\exp(2\pi i \xi/a)$ in 
Eq.~(\ref{7}) takes the two values $\pm 1$ for $\xi=a$ or 
$\xi=a/2$. It has been introduced in Ref.~\cite{Ivchenko1} to 
take into account the symmetry properies of the Bloch functions 
with respect to translation by a single monomolecular layer 
($a/2$) along the $z$ axis.  The Bloch function of the electron 
in the $X_z$ valley changes its sigh under this translation 
whereas the Bloch function of the electron in the $\Gamma$ 
valley does not.  Therefore, the parameter $t_{\Gamma X}$ of 
$\Gamma$--$X$ mixing also should change sigh under such 
translation. This is not important at a plane interface, but it 
must be taken into account when the relative positions of some 
interfaces are considered. We assume 
$|t^{xy,h}_{21}|\ll a^{-1}$:  this is the result of numerical 
calculations.\cite{Ando} 
\end{mathletters}

Unlike Eqs.~(\ref{bcplane}) the boundary conditions 
(\ref{bcrough}) contain the terms depended on $\xi$. They would 
not be important, if $\xi={\rm const}$. Then they relevant to 
the phase shift of the wave functions due to the shift of the 
interface. However, these terms become important when $\xi$ 
depends on $\bbox{\rho}$. Interference     of the electrons 
scattered from the neighboring planes in the vicinity of  
steps (like point 1 on Fig.~1) results in the diffuse  component 
of their wave function. The mean size of the region at the step 
where the interference occurs is the parallel-to-interface 
component of the electron wavelength. Hence the ratio of this 
size to the size of the plane area $l$ characterizes  the 
roughnesses influence on the electrons.

We separate  the diffuse components  
$\varphi_{\Gamma,X_z,X_{xy}}^{l,r}$ of the envelope wave 
functions and write the envelopes as follows: 
\cite{Bass}
\begin{eqnarray} 
\label{10} 
&&\Psi_{\Gamma,X_z,X_{xy}}^{l,r}=\Phi_{\Gamma,X_z,X_{xy}}^{l,r}
+\varphi_{\Gamma,X_z,X_{xy}}^{l,r}, \\
&&\mbox{where} 
\quad \overline{\varphi_{\Gamma,X_z,X_{xy}}^{l,r}}=0.
\nonumber
\end{eqnarray}

\widetext
Using the boundary conditions~(\ref{bcrough}), 
for the envelopes $\Phi_{\Gamma,X_z,X_{xy}}^{l,r}$ and 
$\varphi_{\Gamma,X_z,X_{xy}}^{l,r}$ (see Ref.~\cite{PhysicaE} 
for the details) we obtain
\begin{eqnarray*}
%\label{11}
\Phi_\Gamma^l({\bf r})=T_\Gamma e^{-ip_\Gamma z},\quad 
&\Phi_{X_z,X_{xy}}^l({\bf r})= 
T_{X_z,X_{xy}}e^{\gamma_{X_z,X_{xy}}z},\quad
&\Phi_{X_z,X_{xy}}^r({\bf 
r})=e^{-iqz}+R_{X_z,X_{xy}}e^{iqz},%\nonumber\\
\end{eqnarray*}
\begin{eqnarray*}
\Phi_h^l({\bf r})=e^{ipz}+R_he^{-ipz},\quad%\nonumber\\
&&\Phi_h^r({\bf r})=
T_he^{-\gamma_hz},%\\
\end{eqnarray*}
\begin{eqnarray}
\label{11}
\varphi_{\Gamma}^l({\bf r})=
\frac{2q}{(2\pi)^2}
\int_{-\infty}^{\infty} A_{\Gamma}^l(\bbox{k_\parallel})
\tilde{\xi}(\bbox{k_\parallel})
e^{i(\bbox{\scriptstyle k_\parallel 
\rho}-k_\Gamma z)}\, 
d\bbox{k_\parallel},\nonumber \\ 
\varphi_{X_z,X_{xy}}^l({\bf r})=
\frac{2q}{(2\pi)^2}
\int_{-\infty}^{\infty} A_{X_z,X_{xy}}^l(\bbox{k_\parallel})
\tilde{\xi}(\bbox{k_\parallel})
e^{i\bbox{\scriptstyle k_\parallel 
\rho}+\ae_{\Gamma,X_z,X_{xy}}z}\, d\bbox{k_\parallel},\nonumber 
\\ 
\varphi_{X_z,X_{xy}}^r({\bf r})= \frac{2q}{(2\pi)^2} 
\int_{-\infty}^{\infty} A_{\Gamma,X_z,X_{xy}}^r(\bbox{k_\parallel})
\tilde{\xi}(\bbox{k_\parallel})
e^{i(\bbox{\scriptstyle k_\parallel 
\rho}+k_{X_z,X_{xy}}z)}\,
d\bbox{k_\parallel}, \nonumber\\
\varphi_h^l({\bf r})=
\frac{2p}{(2\pi)^2}
\int_{-\infty}^{\infty} A_h^l(\bbox{k_\parallel})
\tilde{\xi}(\bbox{k_\parallel})
e^{i(\bbox{\scriptstyle k_\parallel 
\rho}-k_hz)}\,
d\bbox{k_\parallel}, \nonumber\\
\varphi_h^r({\bf r})=
\frac{2p}{(2\pi)^2}
\int_{-\infty}^{\infty} A_h^r(\bbox{k_\parallel})
\tilde{\xi}(\bbox{k_\parallel})
e^{i\bbox{\scriptstyle k_\parallel 
\rho}-\ae_hz}\,
d\bbox{k_\parallel}. %\nonumber
\end{eqnarray}
Where
\begin{eqnarray*}
%\label{12}
&&k_\Gamma(\bbox{k_\parallel})=
\sqrt{2m_\Gamma(\varepsilon_e-E_\Gamma^l)-\bbox{k_\parallel}^2},
\quad
\ae_{X_z}(\bbox{k_\parallel})=
\sqrt{2m_{Xl}^l(E_{X_z}^l-\varepsilon_e)+\bbox{k_\parallel}^2},\\
&&\ae_{X_{xy}}(\bbox{k_\parallel})=
\sqrt{2m_{Xt}^l(E_{X_{xy}}^l-\varepsilon_e)+\bbox{k_\parallel}^2},
\quad
k_{X_z}(\bbox{k_\parallel})=
\sqrt{2m_{Xl}^r(\varepsilon_e-E_{X_z}^r)-\bbox{k_\parallel}^2},\\
&&k_{X_{xy}}(\bbox{k_\parallel})=
\sqrt{2m_{Xt}^r(\varepsilon_e-E_{X_{xy}}^r)-\bbox{k_\parallel}^2},
\quad
k_h(\bbox{k_\parallel})=
\sqrt{2m_{h}^l(E_h^l-\varepsilon_h)-\bbox{k_\parallel}^2},\\
&&\ae_h(\bbox{k_\parallel})=
\sqrt{2m_{h}^r(\varepsilon_h-E_h^r)+\bbox{k_\parallel}^2},
\quad
{\rm Im}\,k_{X_z,X_{xy},h}\geq 0,\\
&&T_\Gamma=\frac{2iqt^z_{13}\eta(\xi)}{t^z_{44}\gamma_{X_z}},\;\;\;
T_{X_z}=-\frac{2iq}{t^z_{44}\gamma_{X_z}},\;\;\;
R_{X_z}=-1-\frac{2iq}{t^z_{44}\gamma_{X_z}},
\nonumber\\
&&T_{X_{xy}}=\frac{2iq}{t^{xy}_{21}+t^{xy}_{22}\gamma_{X_{xy}}},\;\;\;
R_{X_{xy}}=-1+\frac{2iq}{t^{xy}_{21}+t^{xy}_{22}\gamma_{X_{xy}}},
\\
&&T_h=-\frac{2ip}{-t^h_{21}+t^h_{22}\gamma_h},\;\;\;
R_h=-1-\frac{2ip}{-t^h_{21}+t^h_{22}\gamma_h},
\nonumber\\
&&A_\Gamma^l=\frac{it^z_{13}\eta(\xi)}{t^z_{44}},\quad
%\nonumber\\
A_{X_z}^r=-\frac{i}{t^z_{44}}
\left(\frac{t_{13}^zt_{41}^z}{\gamma_{X_z}}+ 
1-t^z_{44}\right),\quad
%\nonumber \\
A_{X_z}^l=-\frac{k_{X_z}}{\ae_{X_z}{t^{z}_{44}}^2}
\left(\frac{t_{13}^zt_{41}^z}{\gamma_{X_z}}+ 
1-t^z_{44}\right),
\nonumber \\
&&A_{X_{xy}}^r=i
\frac{t^{xy}_{22}\ae_l\ae_{X_{xy}}(1-t^{xy}_{22})}
{(t^{xy}_{21}+\ae_{X_{xy}}t^{xy}_{22})
(t^{xy}_{22}\ae_l+ik_{X_{xy}})},\qquad
%\nonumber\\
A_{X_{xy}}^l=k_z
\frac{\ae_{X_{xy}}(1-t^{xy}_{22})}
{(t^{xy}_{21}+\ae_{X_{xy}}t^{xy}_{22})
(t^{xy}_{22}\ae_l+ik_{X_{xy}})},
\nonumber\\
&&A_h^r=
-\frac{k_h(1-t^h_{22})}
{t^h_{22}(\ae_ht^h_{22}-t^h_{21})},\qquad
%\nonumber\\
A_h^l=
i\frac{\ae_h(1-t^h_{22})}
{\ae_ht^h_{22}-t^h_{21}}.
\nonumber
\end{eqnarray*}
\narrowtext
Here 
$\tilde{\xi}(\bbox{k_\parallel})=\int\xi(\bbox{\rho})
e^{-i\bbox{k_\parallel \rho}}\,d\bbox{\rho}$, 
$\gamma_{X_z,X_{xy},h}=\ae_{X_z,X_{xy},h}(0)$, 
$p_\Gamma=k_\Gamma(0)$; $E_\Gamma^l$, $E_{X_z}^l$, 
$E_{X_{xy}}^l$, $E_{X_z}^r$, $E_{X_{xy}}^r$, $E_h^l$, and 
$E_h^r$ are energies of extrema of the appropriate bands.  
Integration in Eq.~(\ref{11}) is carried out over the whole 
plane because $\xi(\bbox{\rho})$ is not periodical function of 
$\bbox{\rho}$.  The values of normal-to-interface components of 
the wave vectors of the electrons $q$ and holes $p$  are 
determined by the boundary conditions at the interfaces:  
$z=-d_1$ for $p$ and $z=d_2$ for $q$ (where $d_1$ and $d_2$ are 
widths of GaAs and AlAs layers).  In general, they depends on 
the valley under consideration:  $\tan 
\frac{qd_2}{2}=-\frac{q}{\gamma_{X_z}}$ for  electrons in the 
$X_z$ valley, $\tan 
qd_2=-\frac{2q}{t^{xy}_{22}\gamma_{X_{xy}}+t^{xy}_{21}}$ for  
electrons in the $X_{xy}$ valley, and $\tan 
pd_1=-\frac{2p}{t^h_{22}\gamma_h-t^h_{21}}$ for the holes.  We 
assume, however, the  strong confinement of electrons and holes  
in the appropriate layers $\gamma_{X_z,X_{xy},h} \gg p,q$, so 
that $p\approx \pi/d_1$ and $q\approx \pi/d_2$.

The wave function of the electron in the GaAs $\Gamma$ valley is 
small, $p_\Gamma\ll p$; nevertheless, it is real. This 
distinguish the short-period GaAs/AlAs superlattices from other 
type-II structures, where the electron wave function decays 
rapidly from the interface. The electron density is large in 
AlAs and small, but almost constant, in GaAs. This small part of 
the electron density could be essential for the exiton 
recombination would the effective parameter of $\Gamma$--$X$ 
mixing $t_{13}^z$ be sufficently large.

\subsection{Radiative decay rates of indirect excitons at a rough 
interface}

To determine the wave functions $f_e(z)$ and $f_h(z)$, we have 
to insert the corresponding Bloch amplitudes into expressions 
for the envelopes $\Psi_e$ and $\Psi_h$ (\ref{10}). For 
instance, for the $X_z$ exciton at a plane interface, we have 

\begin{eqnarray}
\label{13}
&&f_e({\bf r})=\frac{1}{\sqrt{N_1}}\left\{
\begin{array}{ll}
T_\Gamma u_\Gamma({\bf r})e^{\gamma_\Gamma z}\\
+T_{X_z}u_{X}({\bf 
r})e^{\left(\gamma_{X_z}-\frac{2\pi i}{a}\right)z},& z<0,\\ \\ 
u_{X}^*({\bf r})e^{i\left(q-\frac{2\pi i}{a}\right)z} \\
+R_{X_z}u_{X}({\bf r})e^{-i\left(q-\frac{2\pi i}{a}\right)z},& 
z>0,\\ 
\end{array} 
\right.\nonumber\\ 
&&\ \\ 
&&f_h({\bf 
r})=\frac{1}{\sqrt{N_2}}\left\{ 
\begin{array}{ll} v({\bf r})e^{ipz}+R_hv^*({\bf 
r})e^{-ipz},& z<0,\\ T_hv({\bf r})e^{-\gamma_hz}, &z>0.  
\end{array}
\right. \nonumber
\end{eqnarray}
Where $N_1$ and $N_2$ are numbers of atoms in the AlAs 
and GaAs layers, $u_{\Gamma}({\bf r})$, $u_{X}({\bf r})$ and 
$v({\bf r})$ are Bloch amplitudes of electrons in the $\Gamma$ 
and $X$ valleys, and the holes; we assume these amplitudes to be 
periodical functions of ${\bf r}$.

At a rough interface we have to add also the diffuse components 
of the wave functions. To do that, we have to multiply  
$\varphi({\bf r})$ (\ref{11}) by the corresponding Bloch 
amplitudes. Usually the mean size of  Bloch amplitudes is 
small in comparison with the lattice constant. This allows us 
to assume that $\bbox{\nabla}$-operator in Eq.~(\ref{2}) acts 
only on these amplitudes and to separate the integration of them 
from integration of the envelopes. Then the matrix element 
(\ref{2}) can be written as ${\cal P}={\cal P}_1+{\cal 
P}_2+{\cal P}_3$, where 

\begin{eqnarray} 
\label{14} 
&&{\cal P}_1=\sum_{\Gamma,X}U_{\Gamma,X}\int 
\Phi_{\Gamma,X_z,X_{xy}} \Phi_h\,dz\,d\bbox{\rho}, \nonumber \\ 
&&{\cal P}_2=\sum_{\Gamma,X}U_{\Gamma,X}
\int [\Phi_{\Gamma,X_z,X_{xy}} 
\varphi_h+ \Phi_h \varphi_{\Gamma,X_z,X_{xy}}] 
\,dz\,d\bbox{\rho},  \\ 
&&{\cal P}_3=\sum_{\Gamma,X}U_{\Gamma,X}
\int \varphi_{\Gamma,X_z,X_{xy}} \varphi_h
\,dz\,d\bbox{\rho}.  \nonumber 
\end{eqnarray}
Here $\Phi=\Phi^l$, $\varphi=\varphi^l$ if $z<\xi$; 
$\Phi=\Phi^r$, $\varphi=\varphi^r$ if $z>\xi$;
$U_\Gamma=\Omega_0^{-1}\int_{\Omega_0} u_\Gamma({\bf r}) 
\bbox{\nabla} v({\bf r})\,d{\bf r}$,
$U_X= \Omega_0^{-1}\int_{\Omega_0}u_{X}({\bf r})
\bbox{\nabla} v({\bf r})\,d{\bf r}$,
and $\Omega_0$ is the unit cell.

The rate of the exciton recombination is  
\begin{eqnarray}
\label{w}
&&w=\Lambda\left(
|{\cal P}_1|^2+{\cal P}_1{\cal P}_2^*+{\cal P}_1^*{\cal P}_2
+{\cal P}_1{\cal P}_3^*+{\cal P}_1^*{\cal P}_3
+|{\cal P}_2|^2\right), \nonumber\\
&&\Lambda=\frac{4\hbar e^2\omega}{3m_e^2c^3}G^2({\bf 0}).
\end{eqnarray}
Where $\hbar\omega$ is the exciton energy, $e$, $m_e$ and $c$ are 
the fundamental constants. 

The luminescence magnitude $I(t)$ is 
proportional to the recombination rate $w$ and the number of 
excitons at the time $t$. We assume this number to be 
proportional to $\exp(-wt)$ (or $\exp[-(w_0+w)t]$, if some 
nonstochastic process with the rate $w_0$ occurs).  The $w$ 
value is stochastical, since it depends on $\xi$.  Therefore, to 
determine the luminescence magnitude, we have to average the 
value of $w\exp(-wt)$ over the realization of the random 
function $\xi$.  This could be done if we know the distribution 
$P(w)$ of the $w$ value:  $\overline{w\exp(-wt)}=\int_0^\infty 
w\exp(-wt)P(w)\,dw$.  The distribution $P(w)$ essentially 
depends on ${\cal P}_1$, whether or not it vanishes. 

If ${\cal P}_1=0$ (i.e., if 
the exciton recombination at a plane interface is forbidden) then 
$w$ is proportional to  squared module of 
${\cal P}_2$. The linear dependence between ${\cal P}_2$ and the 
random variable $\xi$ follows from Eqs.~(\ref{11}) and 
 (\ref{14}).  Therefore, if the distribution of $\xi$ is 
 Gaussian, then the distribution of ${\cal P}_2$ is also 
Gaussian and the distribution of $w$ is exponential.  This means 
applicability of arguments of Refs.~\cite{Klein,Minami}, so that 
$I(t)$ is determined by Eq.~(\ref{1}) where $w_r=\Lambda|{\cal 
P}_2|^2$.  For the case of $X_{xy}$ exciton we have
\begin{eqnarray*}
{\cal P}_2&=&\frac{4pqU_X}{\sqrt{N_1N_2}}
\sum_{\bf g}\left[
\frac{1-t_{22}^h}{\gamma_{X_{xy}}^2} 
\tilde{\xi}\left({\bf q}_X+{\bf g}\right)
\right.\\
&+&\left.\frac{1-t_{22}^{xy}}{\gamma_h^2}
\tilde{\xi}^*\left({\bf q}_X+{\bf g}\right)
\right],
\end{eqnarray*}
and
\begin{equation}
\label{15}
w_r=\frac{16\pi^4a^4|U_X|^2\Lambda}{d_1^3d_2^3}
\left[
\frac{(1-t_{22}^h)^2}{\gamma_{X_{xy}}^4}+\frac{(1-t_{22}^{xy})^2}{\gamma_h^4}
\right]
\tilde{W}\left(\frac{\pi}{a}\right).
\end{equation}
Where $\tilde{W}({\bf k})=
\int W(\bbox{\rho})e^{-ik\rho}\,d^2\bbox{\rho}$
is the Fourier transform of
the correlation function; ${\bf q}_X=\left\{2\pi/a,0,0\right\}$ 
is the wave number of the $X$ valley, ${\bf g}$ is a 
two-dimensional reciprocal lattice vector, it arises here since 
integration in Eq.~(\ref{11}) has not been restricted by the 
first Brillouin zone. 

If ${\cal P}_1\neq 0$ (i.e., if 
the exciton recombination at a plane interface is allowed) then
the linear  with respect to ${\cal P}_2$ terms in Eq.~(\ref{w}) 
are nonzero.  This allows to omit the terms with  ${\cal P}_3$ 
and $|{\cal P}_2|^2$, which are quadratic in $\xi$, or replace 
them with their average values.  Then $w$ becomes the linear 
function of the random variable $\xi$.  If the distribution of 
$\xi$ is Gaussian then the distribution of $w$ is Gaussian too, 
i.e., 
\[ 
P(w)=\frac{1}{\sigma\sqrt{2\pi}}e^{-\frac{(w-\overline{w})^2}{2\sigma^2}}.
\] 
Where $\overline{w}=\Lambda|{\cal P}_1|^2$, and 
$\sigma=\left[\overline{|w-\overline{w}|^2}\right]^{1/2}$.

Hence
\begin{eqnarray}
\label{16}
I(t)&=&\frac{e^{-w_0t}}{\sigma\sqrt{2\pi}}
\int_0^\infty we^{-wt-\frac{(w-\overline{w})^2}{2\sigma^2}}\,dw
\\
&=&e^{-(\overline{w}+w_0)t}
\left[
\frac{\sigma}{\sqrt{2\pi}}+\frac{\overline{w}-\sigma^2t}{2}
e^{\frac{\sigma^2t^2}{2}}\mbox{erfc}\left(\frac{\sigma 
t}{\sqrt{2}}
\right) 
\right]. 
   \nonumber
\end{eqnarray}

If $|{\cal P}_1|\gg |{\cal P}_2|$, then
$\sigma^2\simeq
2\Lambda^2
|{\cal P}_1|^2 \overline{{\cal P}_2{\cal P}_2^*}
$. For the case of $X_z$ exciton we have
\begin{eqnarray}
\label{16.5}
&&{\cal P}_1=
\frac{2a^3}{\sqrt{d_1d_2}}
\left[
4\left(\frac{d_1}{d_2}\right)
\frac{t_{13}^z}{\gamma_{X_z}t_{44}^z}U_\Gamma+
\left(\frac{a}{d_1}\right)
\frac{1}{\gamma_{h}t_{22}^h}U_X
\right],\\
&&{\cal P}_2=
\frac{2a^2}{\sqrt{d_1d_2}}
\left[
8\left(\frac{d_1}{d_2}\right)
\frac{{t_{13}^z}}{{t_{44}^z}}U_\Gamma+\right.\nonumber\\
&&\phantom{{\cal P}_2=}\left.
\frac{2\pi i a}{t_{22}^hd_1}
\left(
\frac{1}{t_{44}^z\gamma_{X_z}d_2}
-\frac{1-t_{22}^h}{t_{22}^h\gamma_hd_1}
\right)U_X
\right]
\eta(\xi)\tilde{\xi}(0),\nonumber
\end{eqnarray}
so that
\begin{eqnarray}
\label{17}
&&\overline{w}=
\frac{4\Lambda a^6}{d_1d_2}
\left[
4\left(\frac{d_1}{d_2}\right)
\frac{t_{13}^z}{\gamma_{X_z}t_{44}^z}U_\Gamma+
\left(\frac{a}{d_1}\right)
\frac{1}{\gamma_{h}t_{22}^h}U_X
\right]^2,\\
&&\sigma^2=
\frac{2\Lambda a^4\overline{w}}{d_1d_2}
\left[
64\left(\frac{d_1}{d_2}\right)^2
\frac{{t_{13}^z}^2}{{t_{44}^z}^2}U_\Gamma^2+
\right.
\nonumber\\
&&\left.\phantom{\sigma^2=}
\left(\frac{2\pi a}{t_{22}^hd_1}\right)^2
\left(
\frac{1}{t_{44}^z\gamma_{X_z}d_2}
-\frac{1-t_{22}^h}{t_{22}^h\gamma_hd_1}
\right)^2U_X^2
\right]
\tilde{W}(0).\nonumber
\end{eqnarray}
The first terms in square brackets can be interpreted as 
a electron conversion from the $X$ valley of AlAs to 
the $\Gamma$ valley of GaAs followed by the elecron-hole 
recombination; they are small, since $t_{13}^z\ll 1$.  The 
second ones are due to indirect electron-hole 
recombination;\cite{PR98} they occur only at the interface and, 
therefore, have a small factor $a/d_{1,2}$. This factor is not 
so small in short-period superlattices where $d_{1,2}$ are as 
large as a few lattice constants. The indirect electron-hole 
recombination prevails in such structures, if $a/d_{1,2}\gg 
t_{13}^z$. We omit the terms that contain both these factors or 
$|1-t_{44}^z|\ll 1$.

The question arises, how small should be  $|{\cal P}_1|$ in 
order to Eq.~(\ref{1}) holds? This is possible if the   
deviation of $|{\cal P}_2|^2$  from its average value in 
Eq.~(\ref{w}) essentially exceeds $|{\cal P}_1{\cal P}_2^*|$, 
i.e., when  $|{\cal P}_1|^2\ll (h/l)^2|{\cal P}_2|^2$ or
\begin{equation}
\label{criterium}
\frac{\overline{w}^2}{\sigma^2}\ll
\frac{h^2}{l^2}.
\end{equation}

\section{Kinetics of exciton luminescence in type-II 
G\symbol{"61}A\symbol{"73}/A\symbol{"6C}A\symbol{"73}
superlattices.  Experiment}

The undoped GaAs/AlAs type-II superlattices used in this study 
were grown by molecular-beam-epitaxy  at $600^o$C on a 
(100)  GaAs substrate. The sample BP205, where the 
$X_z$ excitons were studied, contains 40 periods of 19.8-$\AA$ 
GaAs/25.5-$\AA$ AlAs. The $X_{xy}$ excitons were studied in the 
sample BP354. It contains 25 periods of 25-$\AA$  
GaAs/83.5-$\AA$ AlAs.   

The time-resolved  photoluminescence of $X_z$ excitons was 
excited by a YAG:Nd pulse laser with  wavelength $532\, nm$, the 
pulse duration was $0.15\,\mu s$. The N$_2$ laser with 
wavelength $337\, nm$ and pulse duration $7\,ns$ was used to 
investigate the time-resolved photoluminescence of $X_{xy}$ 
excitons.

The luminescence was analyzed by a double grating 
monochromator equipped with a photomultiplier. Lifetime 
measurements were made by the time correlated single-photon 
counting technique. The samples were immersed in liquid helium. 

Figures 2 and 3 present the experimental results on the exciton 
decay rates in our samples. 
Theoretical curves was derived from Eqs.~(\ref{1}), (\ref{16}).
The  values of parameters $w_0=320\,c^{-1}$, $w_r=0.002\times 
10^6\,c^{-1}$, $\overline{w}=0.1\times 10^6\,c^{-1}$,  and 
$\sigma=0.61\times 10^6\,c^{-1}$ ensure the best fit with the 
experiment.  We see that Eq.~(\ref{1}) fits the experimental 
data for the decay rates of  $X_{xy}$ excitons in the sample 
BP354,  whereas Eq.~(\ref{16}) is more appropriate for $X_{z}$ 
excitons in the sample BP205. Note that the value of $w_0$, 
which is associated with  the phonon-assisted recombination, is 
small in both curves.  That is really the case at a low 
temperature.  Recombination of $X_{xy}$ excitons is considerably 
slower than that of $X_z$ excitons. This means that the 
interfaces in our samples are perfect enough to apply our theory 
for interpretation of the experimental data.

Expressions (\ref{15}) -- (\ref{17}) allow to estimate the 
function $\tilde{W}(k)$ at the points $k=0$ and $k=2\pi/a$.  
This is not sufficient to determine the function. However, it is 
possible to estimate the parameters of the rough interface if we 
restrict ourself to the particular type of the correlation 
function. We  assume the exponential correlation function
\begin{equation}
\label{18}
W(\rho)=h^2\exp(-\rho/l),
\end{equation}
where $l$ 
is the correlation length. This type of the correlation 
function is more appropriate to our model of the rough interface 
(Fig.~1); it allows to construct the two-position distribution, 
so that the distribution of slopes has a $\delta$-singularity, 
i.e., the slope is always zero exept a set of points (like point 
1) with measure zero.\cite{Berry}  
This is impossible for the Gaussian correlation function 
$W(\rho)=h^2\exp(-\rho^2/l^2)$ most frequently employed in 
theoretical discussions.\cite{Kosobukin}   Fourier 
transform of the exponential function    is
\begin{equation} 
\label{18a} 
\tilde{W}(k)= \frac{2\pi h^2l^2}{(1+k^2l^2)^{3/2}}. 
\end{equation} 
Unlike the Gaussian function it has not exponential factor, 
which is small at a large $k$.    This is also due to the 
singular points 1; only in the vicinity of these points the 
momentum relaxation of indirect $X_{xy}$ excitons is possible. 

If we assume that correlation functions are equal for the 
interfaces of both our samples, then substitution of 
Eq.~(\ref{18a}) into Eqs.~(\ref{15}) and (\ref{17}) allows to 
find the values of $h$ and $l$.  
The decay parameters of the wave functions  
$\gamma_{X_z,X_{xy},h}$ in these 
expressions are determined by Eq.~(\ref{11}) for the known 
energies of the electrons or holes.  As regards to 
$U_{\Gamma,X}$, these values can be estimated only from the 
band structure calculations for GaAs and AlAs. However,  the 
first terms in the expression for $\overline{w}$ and $\sigma^2$ 
(\ref{17}) can be omitted.  Indeed, $t^z_{13}=t_{\Gamma 
X}m^{\rm GaAs}_\Gamma/(m_ea\gamma^r)<0.06$, whereas 
$a/d_{1}=2/7$, i.e., the indirect recombination of 
$X_z$ exitons at the interface prevails in  our samples.   Then 
the values of ${\sigma^2}/{\overline{w}^2}$ and 
$w_r/{\overline{w}}$, which are determined from experimental 
data, become independent of $U_{\Gamma,X}$.  For our experiments 
this estimation yields 
\widetext 
\begin{eqnarray} 
\label{19} 
&&\frac{\sigma^2}{\overline{w}^2}=
\frac{1}{2}\left(\frac{2\pi}{a}\right)^2
\left(
\frac{\gamma_h}{\gamma_{X_z}d_2}-
\frac{1-t^h_{22}}{t^h_{22}d_1}
\right)^2\tilde{W}(0),
\nonumber \\
&&\frac{w_r}{\overline{w}}=
\frac{4\pi^4\gamma_h^2{t^h_{22}}^2
{d_1^{X_{xy}}}^3d_2^{X_{xy}}}
{a^4{d_1^{X_z}}^3{d_2^{X_z}}^3}
\left[
\frac{(1-t_{22}^h)^2}{\tilde{\gamma}_{X_{xy}}^4}+
\frac{(1-t_{22}^{xy})^2}{\tilde{\gamma}_h^4}
\right]
\tilde{W}\left(\frac{2\pi}{a}\right).
\end{eqnarray}
Here $d_1^{X_z}$, $d_2^{X_z}$, $d_1^{X_{xy}}$, and 
$d_2^{X_{xy}}$ are widths of GaAs and AlAs layers in the samples 
BP205 ($d_1^{X_z}$, $d_2^{X_z}$) and BP354 ($d_1^{X_{xy}}$,  
$d_2^{X_{xy}}$) where $X_z$ and $X_{xy}$ excitons were studied;
\begin{eqnarray*}
&&\gamma_{X_z}=\frac{1}{\hbar}
\sqrt{2m_{Xl}^{GaAs}
\left[E_X^{GaAs}-E_X^{AlAs}-\frac{\hbar^2}
{2m_{Xl}^{AlAs}(d_2^{X_z})^2} \right]},\\
&&\gamma_h=\frac{1}{\hbar}
\sqrt{2m_{hh}^{AlAs}
\left[E_h^{GaAs}-E_h^{AlAs}-\frac{\hbar^2}
{2m_{hh}^{GaAs}(d_1^{X_z})^2} \right]},
\quad
t_{22}^h=\frac{m_{hh}^{AlAs}}{m_{hh}^{GaAs}},\\
&&\tilde{\gamma}_{X_{xy}}=\frac{1}{\hbar}
\sqrt{2m_{Xt}^{GaAs}
\left[E_X^{GaAs}-E_X^{AlAs}-\frac{\hbar^2}
{2m_{Xt}^{AlAs}(d_2^{X_{xy}})^2} \right]},
\quad
t_{22}^{xy}=\frac{m_{Xt}^{AlAs}}{m_{Xt}^{GaAs}},
\\
&&\tilde{\gamma}_h=\frac{1}{\hbar}
\sqrt{2m_{hh}^{AlAs}
\left[E_h^{GaAs}-E_h^{AlAs}-\frac{\hbar^2}
{2m_{hh}^{GaAs}(d_1^{X_{xy}})^2} \right]},
\quad
\gamma^r=\frac{1}{\hbar}
\sqrt{2m_{\Gamma}^{AlAs}
\left(E_\Gamma^{AlAs}-E_X^{AlAs}\right)};
\end{eqnarray*}                             
\narrowtext
$E_{\Gamma,X,h}^{GaAs,AlAs}$ are positions of the band extrema 
in GaAs and AlAs, $m_{\Gamma}^{GaAs}$ is  effective mass of 
$\Gamma$ valley of GaAs, $m_{Xl,Xt}^{GaAs,AlAs}$ are longitudial 
and transversal effective masses in $X$ valleys of GaAs and 
AlAs, $m_{hh}^{GaAs,AlAs}$ are effective masses of heavy holes 
in GaAs and AlAs, and $m_e$ is mass of a free electron. We 
assume $t_{21}^h\ll\gamma_h\sim 2/a$; this is the result of 
calculations. \cite{Ando} 

Eq.~(\ref{19}) estimates the values of 
 For 
the  height $h$ and diameter $L$ [$L=4l$ for the distribution 
(\ref{18})]\cite{Berry} of the roughnesses we find $h\approx 
1.25a$ and $L\approx 9a$.  This is in agreement with structural 
reseach of the GaAs/AlAs interface where the steps with the 
height $h=a/2$ and the mean length of 40--200$\,\AA$ were 
observed (see Ref.~\cite{Bechstedt} for the review). 

Rough estimation of $h$ and $l$ values also can be done if we 
assume that criterium (\ref{criterium}) holds. This justifies  
Eq.~(\ref{1}) for $X_z$ excitons where $w_0\equiv 
w_0^{X_z}=\Lambda|{\cal P}_1|^2$ and $w_r\equiv 
w_r^{X_z}=\Lambda|{\cal P}_2|^2$. Using Eq.~(\ref{16.5}), we 
find the expressions for $2w_r^{X_z}/w_0^{X_z}$ and 
$w_r/w_0^{X_z}$ [unlike $w_r^{X_z}$ the $w_r$ value  
correspondes to $X_{xy}$ excitons (Fig.~3) and determined by 
Eq.~(\ref{15})]. These expressions accept the form of 
Eq.~(\ref{19}) after the substitutions 
$\sigma^2/\overline{w}^2\rightarrow 2w_r^{X_z}/w_0^{X_z}$ and 
$w_r/\overline{w}\rightarrow w_r/w_0^{X_z}$ of their left sides. 
The values of $w_0^{X_z}=0.11\times 10^6\,c^{-1}$ and 
$w_r^{X_z}=0.38\times 10^6\,c^{-1}$ ensure the best fit of the 
dashed line (Fig.~2) with experiment. This estimation yields 
$h=a$, $L=8.8a$, which are close to the values obtained from 
Eq.~(\ref{16}). For this reason both theoretical curves (Fig.~2) 
fit experimental data at small times.  Nevertheless, 
Eq.~(\ref{16}) better fits the experimental data at large times 
where it unsure the slower decay of the luminescence:  
$I(t)\propto \exp (-\overline{w}t)/t$, instead of $I(t)\propto 
\exp (-w_0t)/t^{3/2}$ as it is predicted by Eq.~(\ref{1}).

\section{Discussion}

In this paper we investigate the exciton luminescence in type II 
GaAs/AlAs superlattices. We use the envelope function 
approximation to consider the exciton recombination at an 
interface. To justify this approach, we have to note that 
envelope function approximation has been used only to find the 
reflection and transmission coefficients.  While the Bloch 
functions $f_e$ and $f_h$ has been used to find the probability 
of the exciton recombination.  The error arises only when we 
consider the Bloch amplidudes $u_\Gamma({\bf r})$, $u_X({\bf 
r})$, and $v({\bf r})$ as periodical functions at the interface.  
Indeed, the deviation of these amplitudes from their bulk values 
arises only at a small distance from the interface; this 
deviation is especially small for the contacts of similar 
materials (e.g., GaAs/AlAs).\cite{Ando,Allmen} 

It seems the boundary conditions  (\ref{bcplane}) connect a 
very few valleys of the electron spectrum to consider the 
interface influence on the exciton recombination; nevertheless, 
it is not the case. Indeed,  the electron wave functions in the 
valleys that are not explicitly involved in Eq.~(\ref{bcplane}) are 
strongly localized at the interface. This allows to consider 
them in terms of the boundary conditions where the parameters 
$t_{ik}$ are influenced by these valleys.  This procedure had 
been described  when Eq.~(\ref{bcplane}) was derived. The error arises 
only when these parameters are considered as independent of the 
electron energy; that is possible if the energy difference 
between the bottoms of the appropriate valleys considerably 
exceeds the exciton energy.  Note that a lot (about 10) of the 
electron bands are sometimes taken into account when the 
parameters of the interface matrix  
are calculated.\cite{Grinyaev}

We use the boundary conditions for the envelope wave function 
to consider $\Gamma-X$ mixing of  electrons at 
the interface.  This approach is more general than the kinetic 
model proposed in Ref.~\cite{Maaref}. The kinetic equation where 
the electron states in the $\Gamma$ and $X$ valleys are 
considered as independent can be used  for low $\Gamma-X$ 
mixing.  Only in that case it is possible to add the 
probabilities for the electron to be in $\Gamma$ and $X$ 
valleys.  It should be noted that we also assume the small value 
of $\Gamma-X$ mixing ($|t_{13}^z|\ll 1)$. However, this 
approximation is not principal for our consideration; it only 
makes the results [Eqs.~(\ref{11}), (\ref{15}), (\ref{17}), and 
(\ref{19})] not so cumbersome.

Influence of a nonstochastic process on the exciton recombination 
in Ref.~\cite{Klein} is considered by the exponential factor 
$e^{-w_0t}$. This factor could be obtained if we 
insert the corresponding term in $\tau$-approximation  
into  the kinetic equation for the exciton density.    If the 
$\tau$-approximation is not applicable for the process, then 
this factor becomes nonexponential.\cite{Krivirotov} Correlation 
between  stochastic and nonstochastic processes changes  the 
second factor in Eq.~(\ref{1}).  In this case the probability of 
the exciton recombination $w$ is not a simple sum of  the 
probabilities of each process.  As the result, the additional 
terms arise  in the expression for $w$ [the second and  third
terms in Eq.~(\ref{w})]. These terms are linear in  the 
 stochastic variable, so that their averages vanish. For this 
 reason they are not important when the mean intensity of the 
 luminescence or the light absorption\cite{PhysicaE} is 
considered. However, they are important for the kinetic 
phenomena, because they determine the mean square of the 
deviation $\sigma$  of the stochastic variable from its mean 
value. The nonexponential behavior of the decay rate 
Eq.~(\ref{16}) valids any time when  linear with respect to 
the stochastic variable terms are main in 
the expression for $w$.   Such a situation can occur also in 
other type-II semiconductor structures where the interface 
influence is essential, e.g., in quantum dots.\cite{Govorov}

Expressions (\ref{15}) and (\ref{17}) relate parameters of the 
radiative decay rates ($w_r, \overline{w}$, and $\sigma$) with 
the correlation function of the rough interface. The values of 
the Fourier transform of this function at two particular points, 
$k=0$ and $k=2\pi/a$, are necessary for these relations. This 
allows to estimate the parameters only for  simplest 
functions [like Eq.~\ref{18}]. The real interface might be more 
complicated. In particular,  a few different 
scales could be characteristic for the roughnesses at the interface. 
Expressions (\ref{15}) and (\ref{17}) takes into account all 
these factors; however, it is impossible to determine more than 
two parameters  from the time-resolved luminescence experiments 
only.

Comparing the experimental results (Figs. 2 and 3), we see  that 
mean lifetime of $X_{xy}$ excitons essentially exceeds that of 
$X_z$ excitons.  This happens due to  
recombination of  $X_z$ excitons  at a plane interface. 
Meanwhile, influence of the roughnesses, i.e., the 
nonexponential factor in $I(t)$, is more essential for $X_z$ 
excitons. This can be understood from our analysis. Indeed, 
$\sigma\propto \tilde{W}(0)$, whereas $w_r\propto 
\tilde{W}(2\pi/a)$ while $\tilde{W}(2\pi/a)\ll \tilde{W}(0)$. 
The recombination occurs in some region near the step (point 1 
in Fig.~1). The size of this region is of the order of 
$|{\bf q}_\parallel|^{-1}$, where ${\bf q}_\parallel$ is the 
parallel-to-interface component of the electron wave vector.  
This region is large for  $X_z$ electrons  ($|{\bf 
q}_\parallel|\simeq r_B^{-1}$, where $r_B$ is the exciton 
radius) but it is small for $X_{xy}$ electrons ($|{\bf 
q}_\parallel|\simeq 2\pi/a$).  As the result, the small factor 
[of the order of $(a/l)^3$] arises in the expression for $w_r$.

In conclusion,   kinetics of the exciton luminescence 
at a rough interface has been considered. The Klein {\it at al.} 
law (\ref{1}) is shown to be valid for the decay rate of 
$X_{xy}$ excitons, whereas the more complicated expression 
(\ref{16}) is applicable for $X_z$ excitons. Expressions  
(\ref{15}) and (\ref{17}), which relate the parameters of the 
exciton kinetics with statistical characteristics of the rough 
interface, allow to estimate some of these characteristics from 
the experimenal data.  The values of the mean height  $7\,\AA$
and length $50\,\AA$ of the roughnesses obtained from 
our experiments are in a good agreement with the results of 
structural investigations of the GaAs/AlAs interface.

 \acknowledgments
Authors  wish to thank  
Prof.\ E.\ L.\ Ivchenko  for valuable discussions.   This work 
was supported by the Russian Foundation for the Basic Research, 
Grants No.~99-02-17019, 98-02-17896, 00-02-17658, and the 
Program ''Physics of Solid State Nanostructures'' by the Russian 
Interdisciplinary Scintific and Technical Council, Grant 
No.~99-1133.

\begin{figure}
\caption{
The model of the rough interface: side view.
}
%\label{}
\end{figure}

\begin{figure}
\caption{
Temporal evolution of the $X_z$ exciton emmision.
Theoretical curves (dashed and  solid  lines) was derived from 
Eqs.~(\ref{1}) and (\ref{16})   respectively. Dotts  show 
the experimental data.  } %\label{} 
\end{figure} 

\begin{figure} 
\caption{
Temporal evolution of the $X_{xy}$ exciton emmision.
Theoretical curve (solid line)  was derived from 
Eq.~(\ref{1}). Dotts  show the 
experimental data. 
}
%\label{}
\end{figure}


\begin{references}
\bibitem{Klein}M.~V.~Klein, M.~D.~Sturge, and E.~Cohen, Phys. 
Rev. B, {\bf 25}, 4331 (1982).

\bibitem{Minami}F.~Minami {\it et al.}, Phys. Rev. B, {\bf 36}, 
2875 (1987).

\bibitem{Krivirotov}I.~N.~Krivorotov {\it et al.}, Phys. Rev. B, 
{\bf 58}, 10687 (1998).

\bibitem{Nagao}S.~Nagao {\it et al.}, J. of Crystal Growth, 
{\bf 175}, 10687 (1997);
B.~A.~Wilson {\it et al.}, Phys.  Rev.  B, {\bf 40}, 1825   
(1989);
E.~Finkman {\it et al.}, J.~Lumin. {\bf 39}, 57 (1987);
B.~A.~Wilson {\it et al.}, J. Vac. Sci. Technol. B {\bf 6}, 1156 
(1988).  



\bibitem{Ivchenko}E.\ L.\ Ivchenko and G.\ E.\ Pikus, {\it 
Superlattices and Other Heterostructures. Symmetry and Optical 
Phenomena}, second ed. (Springer-Verlag, Berlin, 1997).



\bibitem{PhysicaE}L.~Braginsky, Physica E, {\bf 5}, 142 (1999).

\bibitem{Ando}T.~Ando and H.~Akera, 
Phys.  Rev.  B, {\bf 40}, 11619   (1989). 

\bibitem{Lurssen}D.~L\"uer\ss en {\it et al.},  Phys.  Rev.  B, {\bf 59}, 
15862   (1999). 

\bibitem{Bechstedt}F.~Bechstedt, R.~Enderlein, {\it 
Semiconductor Surfaces and Interfaces. Their Atomic and 
Electronic Structures},  (Berlin: Akademie-Verlag,  1988).

\bibitem{Ivchenko1}I.~L.~Aleiner and E.~L.~Ivchenko, Fiz. Tech. 
Poluprovodn. {\bf 27} 594 (1993) [Sov. Phys. Semicond. {\bf 27} 
(1993)]; Y.~Fu, M.~Willander, E.~L.~Ivchenko, and A.~A.~Kiselev, 
Phys. Rev. B, {\bf 47} 13498 (1993). 


\bibitem{Bass}F.\ G.\ Bass and I.\ M.\ Fuks, {\it Wave 
Scattering from Statistically Rough Surfaces}, (Pergamon Press, 
1979).

\bibitem{PR98}L.~Braginsky, Phys.  Rev. B, {\bf 57}, R6870 (1998).


\bibitem{Berry}M.~V.~Berry, Phil. Trans. {\bf A273}, 611 (1973);
J.~M.~Ziman, {\it Models of Disorder. The theoretical physics of 
homogeneously disordered systems}, (Cambrige University Press, 
1979).

\bibitem{Kosobukin}For the exciton problem this approximation 
has been used in V.~A.~Kosobukin, Fiz.  Tverd. Tela 
{\bf 41}, 330 (1999) [ Phys. of Solid States {\bf 41}, XXX 
(1999)]; see also the references therein.



\bibitem{Allmen} P.~von~Allmen,  Phys.  Rev.  B,  {\bf 46}, 15377
(1992).

\bibitem{Maaref}M.~Maaref {\it et al.}, Solid State Commun., 
{\bf 81}, 35 (1992).

\bibitem{Grinyaev}S.~N.~Grinyaev, private communication; see, 
e.g., S.~N.~Grinyaev and G.~F.~Karavaev, Fiz.  Tverd. Tela 
{\bf 42}, 752 (2000) [ Phys. of Solid States {\bf 42}, XXX 
(2000)].

\bibitem{Govorov} A.~O.~Govorov and A.~V.~Chaplik, Sov. Phys. 
JETP 72, 1037 (1991) [Zh. Eks. Teor. Fiz. 99, 1853 (1991)];
A.~V.~Kalameitsev, A.~O.~Govorov, and V.~Kovalev,  JETP Lett.  
68, 669 (1998) [Pis'ma  Zh. Eks. Teor. Fiz. 68, 634 (1998)].  

\end{references}
\end{document}